# Ticketing System: A Descriptive Research on the Use of Ticketing System for Project Management and Issue Tracking in IT Companies

Kent Darryl M. Aglibar
Graduate Studies, University of the East

Garret Christopher T. Alegre
Graduate Studies, University of the East

Gerald I. Del Mundo
Graduate Studies, University of the East

Kenny Francis O. Duro
Graduate Studies, University of the East

Nelson C. Rodelas
Graduate Studies, University of the East
(Corresponding Author)

**Abstract**

*Purpose* – IT companies are popular in the present time as technology arises in the entire world. Incidents and Requests for either clients or company internal are hard to manually track as it is a day-to-day transaction. To keep all the logs, IT Companies are using a ticketing system to track all the records they are gathering. One of the most popular ticketing systems is the JIRA. Either the support team or company internal is benefitting from the use of JIRA as it easily documents all the different transactions and keeps track of different incidents and requests. One of the cons of JIRA is its limited features. The researchers conducted research entitled "Ticketing System: A Descriptive Research on the Use of Ticketing System among Employees of ABC Company" where it aims to determine the effectiveness of JIRA as a ticketing system despite its limited features.

*Method* – The researchers used a descriptive method to analyze the pros and cons of this ticketing system. The researchers also mentioned how this system works to provide knowledge to the audience that is using other types of ticketing systems.

*Conclusion* – It is concluded that this JIRA Ticketing system is helpful for both support groups and internal members of the company as they track all the incidents and request tickets for their clients despite JIRA's limited features.

*Recommendation* – The researchers strongly recommend to venture out and try other analytic models and start with the ones mentioned in this study for a good foundation and idea. This is can provide the research a new set of results that can be useful to both established and startup companies.

*Practical Implication* – This research will benefit not only the established companies but also the start up companies that are currently struggling in choosing an efficient ticketing system that work well not only in ticketing management but also in project management as well.

## INTRODUCTION

Managing day-to-day transactions is one of the most important in project management as this speaks about everyday possible issues. As time goes by, Project Management evolves and it becomes more prominent in all aspects of businesses as it is one of the main keys to sustaining the profit or the market value of the company. As it is one of the most important management teams, companies also adopt social, environmental, economic, and technologies to maintain their value and to remain relevant in competing in the market (McGrath & Kostalova, 2020). Planning is one part of project management and it leads to success if the management plans carefully and implements the plan. In an article by *"Bob Little",* he identified that one of his fellows has a lot more experience in project management and promotes formulas to gain successful projects. The criteria are *"Identify the business case", "Define the Project", "Closing the Project", and "Learn the lessons that the project should teach you".* With these, it can help the newbies in project management to have a guide on how to implement the plan in the project (Little, 2011). At the end of the day, all of us always wanted some tips and guidance to make our projects successful but not all the time projects will go on what we think of it. Some projects fail and it affects the profit of the companies. There are three (3) criteria on how to assess whether the project is successful. First, proper implementation of the plan. It is easy to identify if the project is still on its right track with proper planning. Without a plan, there is no actual tracking and it is more likely no supervision along the way. Second, the value of the project. Companies also need to make sure that the business process flows correctly. In this case, it talks about how the companies and their resources work for the project. One of the most important is to document all the transactions and create an analysis of how the project goes as it is more like a day-to-day transaction. This also speaks about how the ticketing system helps the business process to document and create necessary steps and analysis for planning. Third and last, Client satisfaction. This is all about the business value and at the end of every project the most important is the profit of the company and this will also dictate the company's market value (Pinto & Mantel, 1990).

In information technology, as we are now moving into the digital age, IT Support is engaged in many corporations, institutions, and other organizations to provide technical support, not only limited to computer networks, operating systems, internet connection, computer security, or any other software/hardware issue that may arise (Al-Emran & Chalabi, 2014). Whether your business is small, medium, or large-scale as long as the organization relies on the information system, IT Support is important and play an important role in an organization and its responsibility is to make sure that they control the application or system by safeguarding the data, troubleshooting, and monitoring (Bourne, 2014).

Imagine a scenario where the Philippine stock exchange trading system blackout for a few hours during the peak of trading due to a sudden system update or technical issues such as failure in the database configuration or physical server went down.  This scenario could potentially create ripples across the financial market and possibly be the headline of the news across the country or maybe the world. Fortunately, the chances of such a scenario are quite small because there is a system composed of IT infrastructure management and IT service management that works 24 hours a day, 7 days a week, 365 days a year to provide essential IT support service to attain business needs of such organization. From business-critical service to life-critical service, IT support has an important key role in the digital age (Rai, Jha, & Puvvala, 2015).

Errors might happen even though it's not critical on the daily basis, minor issues arise the common problems are printer & scanner doesn't work, slow internet, computer networks, computer peripherals failure such as faulty keyboard mouse and OS problem such as the blue screen of death troubleshooting could be done remotely or within the infrastructure of the organization (Gohil, & Kumar, 2019). For this reason, the IT support team uses a ticketing tool so that it will easily identify and recognize problems that arise especially if the problem is occurring over time, using the predefined solution the IT support team would quickly escalate the problem of requesting users. With a ticketing system IT support team will be organized and catalog incoming support queries especially if the organization dealing with high volumes of requests from its end-user, besides of that since tickets can prioritize the IT support team can determine which request should be answered first and with that, the support team will increase its efficiency and satisfaction of the end-user (Venkatasubramanian, Suhasini, & Vennila 2021).

ABC Company has been using JIRA for almost 15 years now as it is known for its unique bug tracking which helped the company not only to mitigate hard-to-find issues but also improve the services that the company is providing to its clients. The company is currently using JIRA mainly for tracking deliverables and release testing whenever there's a software update on its existing services or products. The researchers would like to know if JIRA is effective for project management in terms of tracking deliverables and other fields such as issue tracking and release testing.

## METHODOLOGY

The researcher used a Descriptive Method for this specific study. The data comes from different case studies/journals that have been analyzed to provide an analysis. There are many different definitions for data analysis. Somehow, it depends on the author but despite the different definitions, data analysis has a basic goal, to identify the difference and similarities between two objects and provide us our understanding (Pichler & Savenkov, 2010). Case studies are popular in all aspects of research and they are using it as a part of how to gather data. As per Dr. Kenneth Harling of Wilfrid Laurier University, Waterloo, Ontario, Canada, *"A Case study is a holistic inquiry that investigates a contemporary phenomenon within its natural setting"* (Harling, 2012). The researcher used different case studies to analyze and collect data to support this research for the JIRA ticketing system.

### INFORMATION GATHERING

The researcher will gather information from other research studies, which will help the researchers to determine if JIRA is effective for project management in terms of tracking deliverables.

Information will be gathered from the fields of the Information Technology industry like support, data analytics, developers, and other fields. that the researchers would see the various perspectives and angles. With that, researchers will examine the information gathered so that an in-depth understanding of the effect of the ticketing system concerning the project management process, monitoring, and control phase.

## RESEARCH JOURNAL FILTERING

Researchers filtered each journal found and only focused on the analysis of JIRA and its functionality. The following items were considered during the filtering process:

1. Analytics model being used and
2. Data analysis on the functionalities of JIRA
3. Data analysis tool used to execute the analytics model

After the filtering process, the researchers extracted the important details such as; data analysis, Test data, testing method, results, and discussion.

According to Walker for a ticketing system to be good, effective, and efficient, one must have the following features. It must have the capability to collaborate and support all multichannel operations that are available to the customer. It must highlight convenience at all times and issues must be addressed quickly. It also needs to generate a detailed report and analysis. Reports and analyses serve as valuable references on how you can improve your services and maximize the manpower you have for client support (Venkatasubramanian et al., 2021).

Now to measure the effectiveness and efficiency of a ticketing system, there are some metrics or KPIs we can use to measure one as enumerated in LiveVox Media. For instance, all the total number of tickets and their resolution time can be used for reports and analysis. Another is ticket volume and cost per ticket. This determines where customer engagement is happening by looking at where tickets are originating from and how much customer engagement costs you given your current systems (Gohil, & Kumar, 2019).

Lastly, there are reasons why you need an efficient ticketing system. One is that manual tracking procedures are inefficient. Without an automated ticketing system, IT staff must manually assign tasks based on expertise and track the status of the request in spreadsheets. Another is ticket resolution techniques create delays. Manually translating raw technical data into reports that make sense to management is time-consuming and hinders productivity (Kumar, 2015).

## REVIEW OF RELATED LITERATURE

### WHAT IS JIRA?

JIRA is a bug tracking system created by Atlassian. In 2002. It is widely used for issue tracking in software, its advanced customization capabilities make it ideal for work orders, help desks, and other sorts of ticketing systems and so on), as well as project management. JIRA is a mature, sophisticated software for local personalization to match the requirements of a certain project. It has custom fields, issue types, workflows, notifications, as well as user entry screens (Aoyama, 1995).

### JIRA TICKETING SYSTEM

Ticketing System is popular with the majority of IT companies. It is a tool that supports all internal and external transactions. For example, customers just need to keep the ticket number. When they escalate it to the vendors, it is easy to identify what the transaction is because the ticketing system provides tracking, monitoring, and documentation. In other words, it connects vendors and clients remotely (Gohil, & Kumar, 2019).
JIRA is one of the most popular ticketing systems that has been used by many IT companies. Way back in 2002, Atlassian Corporation created this JIRA Ticketing System to monitor and keep track of bugs-related issues. Later on, because of JIRA's advanced features, it has been suitable in all types of transactions including the IT workforce and for project management (Venkatasubramanian et al., 2021).

In 2006, NIF ICCS used JIRA as a tool to keep track of their software development projects. Before using JIRA, NIF ICCS already used different ticketing systems as their tracking but unfortunately, because of its limited features, it encourages them to try JIRA. Using it, their team manages to complete their transaction successfully. Today, NIF ICCS already partnered with JIRA for tracking purposes including work orders and project management-related records (Venkatasubramanian et al., 2021).

In research by *"Hitesh Mohapatra,"* you can create records in JIRA. As part of its features, users that have access (support/management team) can create different records such as Incident, Request, Problem, and Change. Using the ticket number, it is easy to find references especially if the encountered issues are re-occurrence. Aside from that, management can also backtrack tickets to create analysis regarding the quality of their service to clients (Gohil & Kumar, 2019).

### HOW DOES IT WORK?

When a JIRA ticket is raised, the software project and issue type are specified by the reporter. It could be an improvement, a change, or a problem. Pichler & Savenkov, 2010). The initial fields are entered during the ticket creation process, see the table below (Table 1):

Table 1: JIRA Fields Used by Reporter

| Field Name | Description |
| --- | --- |
| Summary | A one-line description of the request |
| Priority | Urgent, Important, Normal, or Low. Urgent issues may be handled as patch releases |
| Component | The product within the project, chosen from a project-specific list |
| Category | Software, Operational Data, Infrastructure, Documentation. |
| Description | A freeform text field describing the request |
| Affects Versions | What software version this request relates to |
| Environment | Where the issue manifests (main facility, side lab, etc.) |
| Origin | Design Review, Coding, Developer Unit Test, Operations, Offline Tests, etc. |
| Reporter | Who requested the change (auto filled) |
| Recommend-ation | If a particular fix is needed, it can be specified here |
| Locos # | A ticket reference to the NIF Operations Problem Log system (not JIRA-based) |
| Wrap-around | A flag indicating that this was a data change that originated in the production environment. |

### *FOCUS POINTS OF JIRA AS A TICKETING SYSTEM*

All JIRA tickets undergo several quality assurances phases:

- Design Checks
- Code Checks
- Desk Reviews
- Quality Control
- Software Testing

Whether or not (and when) each of these phases occurs depends on the JIRA issues' nature (Little, 2011).

### *JIRA'S BASIC PROJECT SETUP*

JIRA can create project-based components from defining the project stakeholders, notifications scheme, permission scheme, user groups, issue type scheme, etc. Project stakeholders can control the information being relayed from one department to another which includes the filtering of issue types to be assigned in each department. With each issue type, workflow schemes are also present where the invoice workflow, desired status types, and workflow transitions are configured (Kumar, 2015).

**LIMITATIONS AND WORKAROUNDS OF JIRA**

**Field-level permissions**

JIRA cannot control the editing of specific fields in a JIRA Ticket. If a field happens to be available on an entry screen, then any user may rectify the chosen field given that the user has permission to alter the JIRA ticket (Gohil, & Kumar, 2019).

*Workload and Release Testing*

ICCS produces about five major releases every year. Each release contains an average of about 200 software change requests. Patches are being deployed on a more frequent basis to fix critical issues or bugs which makes it hard for detailed tracing of the enhancement and product testing. JIRA is only able to provide tools for "current" status, which makes the ability to navigate through various graphs into a specific workload detail hard and is currently beyond the functionalities of JIRA.

**Helpful features of JIRA**

There are 3 features in JIRA that helps the employees in working on the tickets, these are the following:

- Reporter
- Assignee
- Watchers

The Reporter is the one that created the JIRA ticket along with the request needed. The Assignee is the one that works on the ticket and does what needs to be done. The Watchers can be a manager, executive, or another project stakeholder with high authority. The research was done regarding these three features which showed how it can influence the ticket resolution time in a certain project or a specific issue.

| Bin | Completion Time (days) |
|-----|------------------------|
| Day | <=1 |
| Week | > 1 and <= 7 |
| Month | > 7 and <= 31 |
| Year | > 31 and <= 365 |
| Long Term | > 365 |

*Figure 1.* Data Table of Issue Completion Time Bins

SVM using RBF Kernel and the XGBoost tool made the analysis possible. A 20% - 80% data split is used for data testing and training. JIRA ticket completion time is predicted using two binning methods: the quick/slow approach and the novel binning approach that consists of the information found in Figure 1 (Kumar, 2015).

| Bucket | Precision (%) | Recall (%) | F1 (%) |
|--------|---------------|------------|--------|
| Fast | 95 | 91 | 93 |
| Slow | 85 | 92 | 88 |

*Figure 2.* Data table predicting Fast and Slow issue completion time using SVM tool

Figure 2 shows the predicting ticket completion time using the quick/slow bins with the use of SVM together with an RBF Kernel. It shows high recall and precision (Kumar, 2015).  The results are also the same with Figure 3 even with the use of the XGBoost tool.

| Bucket | Precision (%) | Recall (%) | F1 (%) |
|--------|---------------|------------|--------|
| Fast | 95 | 84 | 89 |
| Slow | 77 | 92 | 84 |

*Figure 3.* Data table predicting Fast and Slow issue completion time using XGBoost tool

| Class | Weight (%) |
|-------|------------|
| Assignee | 23 |
| Reporter | 55 |
| Mood | 3 |
| Watchers | 18 |

*Figure 4.* Data table Predicting Fine-grained Binned Issue Completion Time using XGBoost

Figure 4 shows the most weighted features which are shown in the table. These features are essential in the JIRA ticketing system and are being raised by users of the issue without writing any notes or comments as it doesn't play any part in predicting the completion time of a specific issue (Kumar, 2015).

## RESULTS AND DISCUSSIONS

In terms of ticket management, the functionality of JIRA is quite helpful as stated by Solokov "It is reasonable to assume that an assignee will complete various issues in the same manner as their ability to perform their duties is unlikely to change significantly. It's fascinating to see that the number of people watching an issue is related to the time it takes to complete the issue, Priority and resolution. This could imply that the greater the number of people is linked on an issue, the more diligently the assignee will work on that project to completion." (McGrath & Kostalova, 2020).

Having those features boosts the progress in each JIRA Ticket being raised and worked on. May it be for Release Testing, Code Review, Quality Control, etc. The most influential feature, reporter, was given far more weight than any other characteristic; it demonstrates that the person who raises an issue has a strong influence on the matter. This person is usually a project manager or a senior executive. The developer or someone in a position of authority in a company is directly responsible for determining the priority of an issue and has the authority to push it through. Certain tasks must be completed by developers/assignees (McGrath & Kostalova, 2020).

## CONCLUSIONS AND RECOMMENDATIONS

Given that the features of JIRA are helpful enough in managing tickets in the system, it is safe to say that it can be applied to project deliverables having those features can help a particular project to move quickly as it highly influences the ticket assignee to work on it and get the job done on time. Therefore, it can be concluded that this JIRA Ticketing System is helpful for both support groups and internal members of the company as they track all the incidents and request tickets for their clients despite JIRA's limited features.